\documentclass[conference]{IEEEtran}
\IEEEoverridecommandlockouts
\usepackage{cite}
\usepackage{amsmath,amssymb,amsfonts}
\usepackage{algorithmic}
\usepackage{graphicx}
\usepackage{textcomp}
\usepackage{xcolor}
\usepackage{setspace}

\usepackage{amsmath}
\usepackage{mathtools}

\usepackage{caption,subcaption}
\usepackage{color}

\usepackage{subcaption}
\newcommand\numberthis{\addtocounter{equation}{1}\tag{\theequation}}
\usepackage[font=small,labelfont=bf]{caption}

\def\BibTeX{{\rm B\kern-.05em{\sc i\kern-.025em b}\kern-.08em
    T\kern-.1667em\lower.7ex\hbox{E}\kern-.125emX}}
\begin{document}

\title{INTERPLAY: An Intelligent Model for Predicting Performance Degradation due to Multi-cache Way-disabling \vspace{-0.5cm}\\
}

\author{
   \IEEEauthorblockN
    {
		Panagiota Nikolaou\IEEEauthorrefmark{1}\IEEEauthorrefmark{2},
		Yiannakis Sazeides\IEEEauthorrefmark{2},
		Maria K. Michael\IEEEauthorrefmark{1}\IEEEauthorrefmark{2}
		\\ 
    }
  \IEEEauthorblockA
   {
        \IEEEauthorrefmark{1}Department of Electrical and Computer Engineering and KIOS Centre of Excellence, 
        \IEEEauthorrefmark{2}University of Cyprus \vspace{-2cm}
   }
}


\maketitle

\begin{abstract}
Modern and future processors need to remain functionally correct in the presence of permanent faults to sustain scaling benefits and limit field returns. 
This paper presents a combined analytical and microarchitectural simulation-based framework called INTERPLAY, that can rapidly predict, at design-time, the performance degradation expected from a processor employing way-disabling to handle permanent faults in caches while in-the-field. 
The proposed model can predict a program's performance with an accuracy of up to 98.40\% for a processor with a two-level cache hierarchy, when multiple caches suffer from faults and need to disable one or more of their ways.
INTERPLAY is 9.2x faster than an exhaustive simulation approach since it only needs the  training simulation runs for the single-cache way-disabling configurations to predict the performance for any multi-cache way-disabling configuration.



\end{abstract}

\vspace{-0.7cm}
\begin{spacing}{0.996}
\section{Introduction}
Continued device miniaturization has enabled the integration of many cores and larger caches in processors. A modern processor contains multiple caches that take a large fraction of the total chip area (40\%-60\%) \cite{aga2017compute}.  
However, the scaling beneﬁts for caches are confronted with reliability challenges caused by 
dynamic variations, e.g., aging \cite{bowman2009circuit, ottavi2014dependable},
and operation at low voltage levels\cite{wilkerson2008trading}.
To ensure reliable cache operation one can use spares to replace unreliable cache parts \cite{koren2020fault}, however, this 
incurs high area costs~\cite{nikolaou2015modeling}.

One way industry limits spare overheads is with in-the-field mechanisms for disabling cache segments that are detected to suffer from permanent (frequently repeating) faults\cite{intel, chang200765, IBM, Fujitsu}.  
 However, this can degrade performance due to the extra misses caused by the smaller cache capacity \cite{foutris2013measuring}. Therefore, it has become essential to assess the performance degradation caused by different cache-disabling configurations. 

Such analysis can be done at design-time using simulation to determine field return policies (e.g., which cache-way disabling configurations should flag a field return), so that customers do not suffer from large or unknown performance degradation in-the-field. Thus, there is a need to quantify at design-time the performance impact when operating with different cache-disabling configurations to develop field return policies and to inform customers of the expected performance penalty when way-disabling is used. 

\begin{figure}[t]
    \centering
    \includegraphics[scale=0.24]{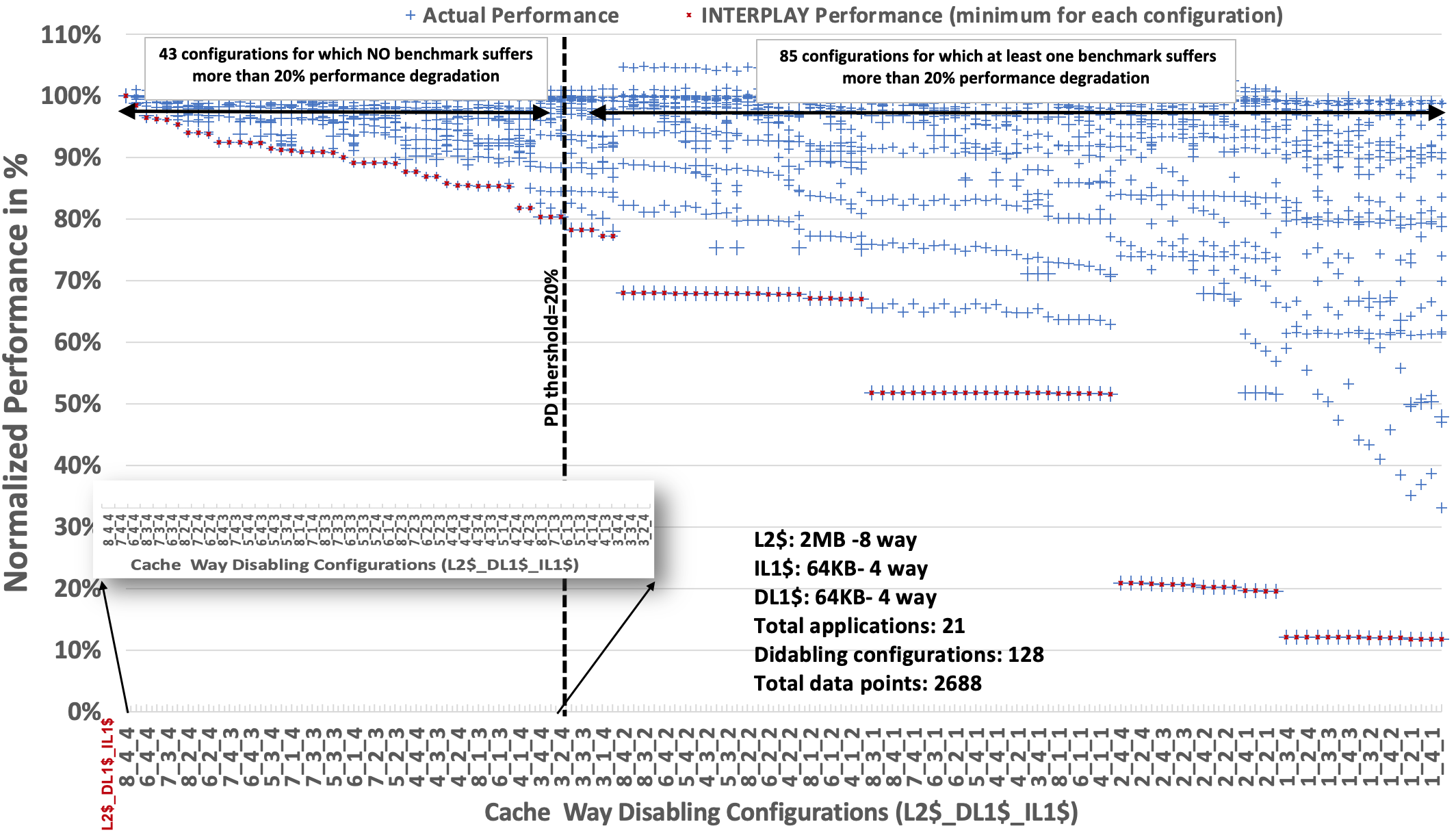}
    \vspace{-0.5cm}
    \caption{Performance Impact for each cache way-disabling configuration (x-axis does not show all points to be legible, inset shows x-axis for configurations for which no benchmark incurs more than 20\% performance degradation).}
    \label{fig:PD}
    \vspace{-0.7cm}
\end{figure}

Cache-disabling can be applied at different granularities, by disabling the line or the entire cache-way that contains a fault~\cite{ patterson1983architecture,intel,chang200765, IBM, Fujitsu}. 
Figure \ref{fig:PD} shows a performance analysis of cache way-disabling for a processor with a two-level cache hierarchy, 
for 21 different applications and all possible cache-way disabling configurations. The processor has a 4-way IL1\$ cache, a 4-way DL1\$ and an 8-way L2\$. Thus, the total number of way-disabling cache configurations of the processor are $4*4*8 = 128$ (assuming there is at least one operational way in each cache) and are shown on the x-axis (denoted by L2\$\_DL1\$\_IL1\$). Configuration $8\_4\_4$ is the baseline configuration with no disabling in any of the caches. Each configuration is evaluated for each of the 21 workloads (i.e., total $128 *21=2688$ data points, the methodology details are given in Section \ref{setup}). The x-axis is sorted according the degradation suffered by any benchmark for a given way-disabling configuration.

It can be seen from Figure~\ref{fig:PD} that for 85 out of 128 (or 66.4\%) of the configurations there is at least one benchmark that exceeds a hypothetical threshold of performance degradation of 20\% (set by the manufacturer based on user requirements). When these way-disabling configurations occur in the field the processor can raise a flag for replacement~\cite{Fujitsu,intel, IBM}. 
One way to determine at design-time the performance degradation due to cache way-disabling, is to exhaustively evaluate, using micro-architectural simulators, all the possible combinations of way-disabling in one or more caches. However, this is, in general, non-practical as the time complexity grows as a product of the number of ways in the cache hierarchy. For example, for a processor with an 8-way IL1\$, 8-way DL1\$, 16-way L2\$ and 20-way L3\$,  20480 simulations are needed per application to cover all cache way-disabling configurations $(8*8*16*20)$. 
An alternative way, is to use analytical methods, in addition to simulation, to predict the performance degradation due to cache disabling. Unfortunately, most existing works predict performance degradation due to disabling in only a single cache\cite{hardy2012performance, hardy2015static, abella2005iatac, yang2002exploiting}. 
Another related work~\cite{choi2017multi}, attempts to reduce power, given a performance constraint, using multi-cache way-power-gating. The work in~\cite{choi2017multi} uses a greedy method to determine the best power-gating configuration. However, the method is only applicable during operation, it is heuristic-based and can suffer from a large performance degradation 
which needs to be detected and remedied on-line. 
In contrast to prior work, our paper aims to predict the performance degradation of all multi-cache way-disabling configurations at design-time in an accurate manner which requires, as we show later, to account for the interplay between caches at different levels.

 


This work proposes INTERPLAY, the first to our knowledge, efficient, design-time, framework capable of predicting in-the-field processor performance degradation due to multi-cache way-disabling.
In particular, our main contributions are:
\begin{itemize}
\item The INTERPLAY framework that 
predicts the performance of any multi-cache way-disabling configuration 
based on a new performance-degradation analytical model that uses microarchitectural statistics from the simulation of single-cache way-disabling training runs. The number of simulations needed for each benchmark by INTERPLAY is proportional to the sum of the number of ways in the various caches instead of their product that is required by the exhaustive simulation approach.
\item We validate INTERPLAY for a specific processor using 14 single-cache way-disabling combinations for training to predict the performance degradation of 114 multi-cache way-disabling configurations, for 21 benchmarks. The results show an average performance degradation prediction error of just 1.4\% with more than 9x reduction in simulation time as compared to the exhaustive simulation approach.


\end{itemize}

The rest of the paper is organized as follows: Section~\ref{framework} describes the INTERPLAY framework. Section \ref{method} presents the proposed analytical prediction model. Section \ref{setup} describes the evaluation setup, and Section \ref{result} presents and discusses the results. Finally, Section \ref{conclusion} concludes this paper and discusses some future directions.

\section{INTERPLAY High Level Flow and Use} \label{framework}


INTERPLAY combines simulation and an analytical model to quickly assess the performance of any multi-cache way-disabling configuration of a processor.
\begin{figure}[t]
    \centering
    \includegraphics[scale=0.45]{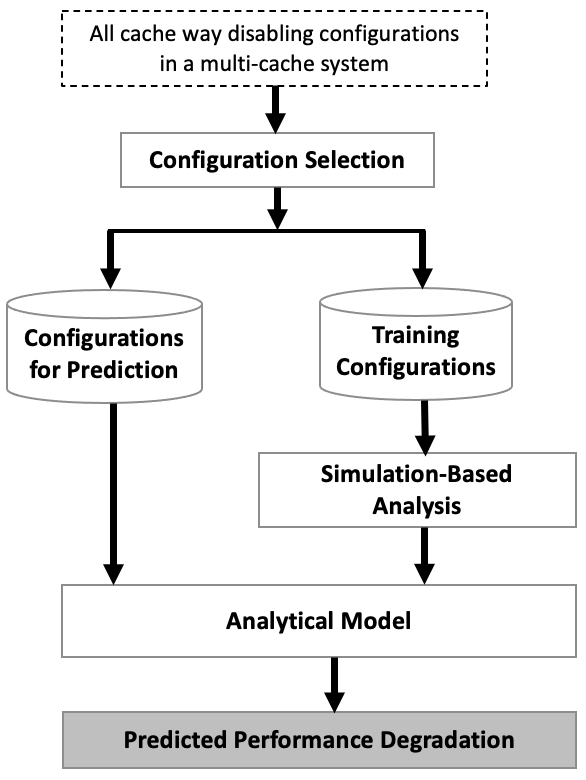}
    \caption{INTERPLAY framework for performance degradation prediction  due to multi-cache way-disabling}
    \label{fig:frame}
    \vspace{-0.7cm}
\end{figure}
It is based on an intelligent selection of a small subset of cache way-disabling configurations that are simulated for an application and used as training dataset to derive the values for the microarchitectural parameters (e.g., cache misses) which are then used by an analytical model to predict the performance of the application for any of the remaining possible way-disabling configurations. Thus, this approach can reduce drastically the number of time-consuming simulations. 

An overview of the proposed framework is given in Figure \ref{fig:frame}. A selection step divides the configurations into, i) those to will be simulated and used as training dataset to feed the analytical model, and ii) those to be predicted using the analytical model. The training configurations are simulated to collect different micro-architectural statistics that feed into the analytical-model which determines the performance impact for the remaining way-disabling configurations (predicted configurations). The computational benefits of INTERPLAY depend on the size of its training set. As we show next, it can be quite small which helps to drastically reduce the time spent on simulations. 

At design-time, a designer can use INTERPLAY to quickly produce a similar analysis to that in Figure~\ref{fig:PD}, to determine 
which way-disabling configurations can cause unacceptable large performance degradation (PD) and should flag a field return. The PD threshold can be defined by the designer, based on the application and its requirements. The effectiveness of INTERPLAY is shown in Figure~~\ref{fig:PD} with red points that indicate the largest degradation predicted by INTERPLAY per way-disabling configuration: it is virtually an exact match with the actual simulation results. Below we explain the workings of INTERPLAY.
\section{Analytical Model for Performance Degradation Prediction} \label{method}

\begin{figure}[t]
    \centering
    \includegraphics[scale=0.43]{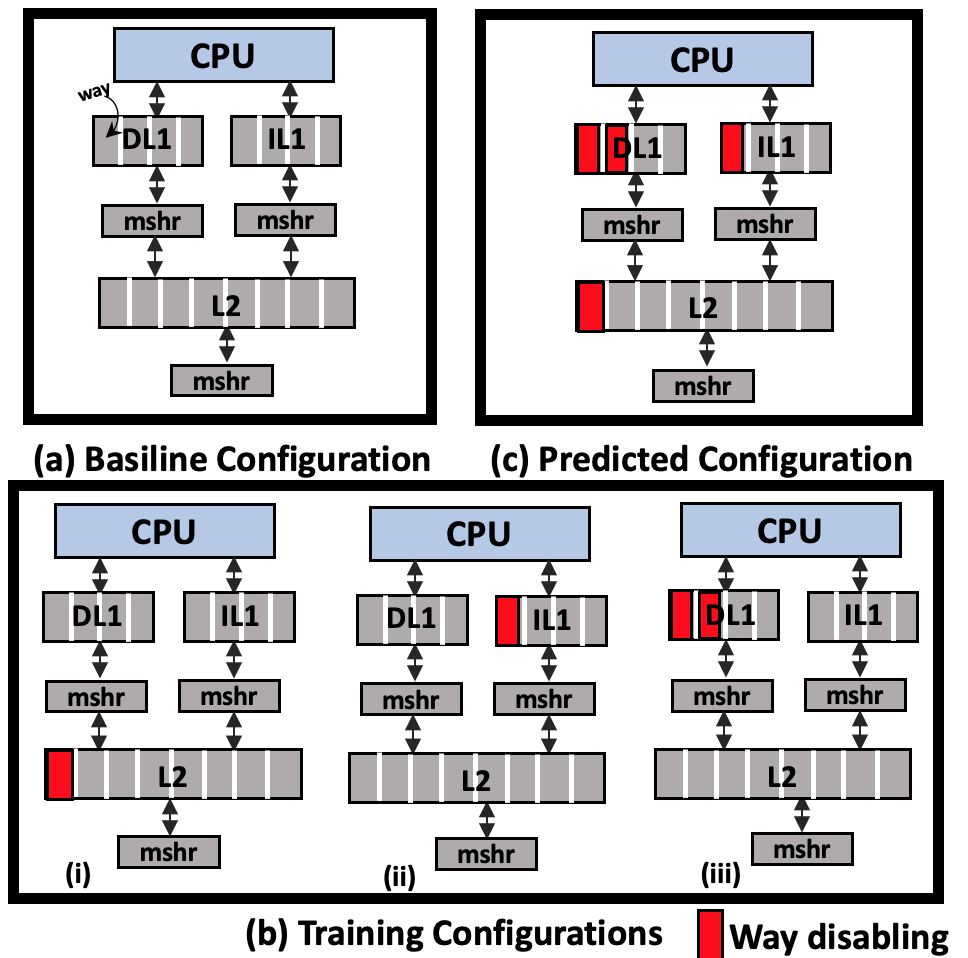}
    \caption{Example of cache hierarchy for (a) baseline configuration (no cache disabling), (b) training configurations with single-way cache-disabling, (c) predicted configuration with multi-way cache-disabling}
    \label{fig:exampleTrainPredict}
    \vspace{-0.48cm}
\end{figure}

This Section describes the models used by INTERPLAY to predict the performance degradation due to multi-cache way-disabling. An example cache hierarchy used to describe INTERPLAY's models is shown in Figure \ref{fig:exampleTrainPredict}(a). We use this example as a point of reference to present the proposed model, which can be generalized for other configurations and types of caches. In this specific case, the hierarchy has two-levels, the level-one or high-level caches consisting of a 4-way instruction cache (IL1\$) and a 4-way data cache (DL1\$), and the low-level cache, in this case, an 8-way unified cache (L2\$). All caches are connected with their representative Miss Status Handling Register (mshr) to track outstanding cache misses.
\subsection{Training Configurations Selection}

One of the main contributions of this paper is the appropriate selection of the training configurations. We keep the set of training configurations small, to reduce the time complexity, but yet sufficient to predict accurately the performance degradation for all the remaining configurations. We select for training the baseline configuration (i.e., without any fault and disabled ways) and all the single-cache way-disabling configurations (i.e., with fault(s) in a single cache). All the remaining multi-cache way-disabling configurations can be predicted by our model. To illustrate, consider the processor in Figure~\ref{fig:exampleTrainPredict}(a) and assume we want to predict the performance of a multi-cache way-disabling configuration with 3-way IL1\$, 2-way DL1\$ and 7-way L2\$ shown in Figure \ref{fig:exampleTrainPredict}(c). Our model will use statistics from the baseline simulation 
(Figure \ref{fig:exampleTrainPredict}(a)) 
plus the three corresponding single-cache way-disabling simulations (Figure \ref{fig:exampleTrainPredict}(b)), each having the same number of ways disabled as in the multi-cache way-disabling configuration but for a single-cache at a time i.e., (i) 4-way IL1\$, 4-way DL1\$ and \emph{7-way L2\$}, (ii) \emph{3-way IL1\$}, 4-way DL1\$ and 8-way L2\$ and (iii) 4-way IL1\$, \emph{2-way DL1\$} and 8-way L2\$.


As the example shows, INTERPLAY attempts to predict the performance of a multi-cache way-disabling in three caches, using the baseline configuration plus the three single-cache way-disabling configurations with the corresponding disabled ways for each individual cache. Consequently, the total number of training configurations that are needed to predict ANY multi-cache way-disabling configuration is equal to 
$1+\sum{(\#ways\ in \ cache_{i}-1)}, \forall{\ cache_{i}}$.
Here, we assume that a functional processor needs to have at least one-way  operational per cache. In an exhaustive simulation approach, the number of cache disabling configurations needed to be simulated in a multi-cache is equal to  $\prod{(\#ways\ in \ cache_{i})}, \forall{ \ cache_{i}}$. Hence, our proposed framework can drastically reduce  the simulation time.
Note that, the exact location of disabled way(s) in a cache is irrelevant, only the \# of ways disabled needs to be considered.

  \begin{table}
\centering
\caption{Parameters used in Performance Degradation Model}
\begin{scriptsize}
\begin{tabular}{|p{1.2cm}|p{7cm}|} 

\hline
\textbf{\textbf{\textbf{\textbf{Parameters}}}} & \textbf{Description}  \\ \hline
$CPI_{P}$    & Predicted CPI \\ \hline   
$CPI_{L}$  & CPI linear model  \\ \hline
$CPI_{EM}$  & CPI model for extra misses in lower-level caches  \\ \hline
$C_{L}$  & Cycles for linear model\\ \hline
$C_{j}$  & Cycles for a configuration \textit{j}, where \textit{j} is:\\ 
& $B$: Baseline configuration (without disabling) \\
& $DT$: DL1\$ Training configuration \\ 
& $IT$: IL1\$ Training configuration\\ 
& $L2T$: L2\$ Training configuration\\  \hline
$EM_{i}$  & Extra misses for cache \textit{i}, where \textit{i} is: \\
&$D\$$: Data cache (DL1\$) \\
&$I\$$: Instruction cache (IL1\$)  \\ 
&$L2\$$: L2 cache (L2\$)  \\ \hline
$M_{i}$  & Misses for cache \textit{i} ($D\$$, $I\$$, $L2\$$) \\ \hline
$M_{i,j}$  & Misses for cache \textit{i} ($D\$$, $I\$$, $L2\$$) in configuration \textit{j} ($B$, $DT$, $IT$, $L2T$) \\ \hline
$M_{i{\rightarrow}m}$  & Misses for cache \textit{i} ($D\$$, $I\$$, $L2\$$) derived from cache \textit{m} ($D\$$, $I\$$, $L2\$$) accesses   \\ \hline

$M_{i{\rightarrow}m,j}$  & Misses for cache \textit{i} ($D\$$, $I\$$, $L2\$$) derived from cache \textit{m} ($D\$$, $I\$$, $L2\$$) for configuration \textit{j} (($B$, $DT$, $IT$, $L2T$))    \\ \hline
$TM_{i}$ &Total Misses for cache \textit{i} ($D\$$, $I\$$, $L2\$$)   \\ \hline
$MR_{i{\rightarrow}m}$  & Miss Rate for cache \textit{i} ($D\$$, $I\$$, $L2\$$) derived from cache \textit{m} ($D\$$, $I\$$, $L2\$$) accesses   \\ \hline
$Penalty_{i}$  & Per miss penalty in cycles for cache \textit{i} ($D\$$, $I\$$, $L2\$$) \\ \hline

\end{tabular}
\end{scriptsize}
\label{tab:parameters}
\vspace{-0.7cm}
\end{table}

\subsection{Analytical Prediction Model}
The selected training configurations are simulated for each benchmark to provide various micro-architectural statistics, such as cycles per instruction (CPI), cache misses and cache accesses, all used as input to the analytical model. Specifically, to formulate the CPI for each predicted configuration (multi-cache way-disabling configuration), we gather and use the statistics from their corresponding single-cache way-disabling training configurations and the baseline configuration.
To determine the predicted CPI we combine, i) a linear CPI model $(CPI_{L})$ that sums the performance degradation, due to the extra cache misses, from all single-cache way-disabling configurations and ii) a CPI model $(CPI_{EM})$ that captures the degradation due to the extra cache misses in the lower-level cache (e.g L2\$) as a result of the interplay between lower and higher level caches.
We define the predicted CPI $(CPI_{P})$ for a multi-cache way-disabling configuration as:
\begin{equation}
  CPI_{P} =CPI_{L} + CPI_{EM}
\end{equation}
%
All the parameters used in the formulation throughout this Section are defined in Table \ref{tab:parameters}. Moreover, for simplicity, when we refer to misses or extra misses, we mean mshr--misses or extra mshr--misses, respectively.\\
\emph{\bf{Linear CPI model ($CPI_{L}$):}}
The first objective, is to estimate $CPI_{L}$. Since the number of committed instructions of a benchmark run is the same for all the configurations, we  replace $CPI$ with cycles $C$. Hence, we use:
\begin{align*}
  C_{L}&=(C_{DT}- C_{B})+(C_{IT}- C_{B})+\\
   &(C_{L2T}- C_{B})+ C_{B} \numberthis 
   \label{eq:model}
\end{align*}
This equation sums the extra cycles contributed from the three single-cache way-disabling training configurations $(C_{DT}, C_{IT}$ and $C_{L2T})$. To this end, we subtract from each training configuration cycles the cycles of the baseline configuration ($C_{B}$). At the end, we add to these differences the cycles of the baseline configuration to estimate the total cycles. 
For example, if we want to predict the CPI for a disabled L2\$ with 1 remaining way, a DL1\$ with 2 remaining ways and an IL1\$ with 3 remaining ways, assuming a baseline configuration of L2\$=8 ways, DL1\$=4 ways and IL1\$=4 ways, then  equation(\ref{eq:model}) becomes:
\begin{align*} 
  C_{L}&=(C_{8\_2\_4}- C_{8\_4\_4})+(C_{8\_4\_3}- C_{8\_4\_4})+\\
  & (C_{1\_4\_4}- C_{8\_4\_4})+C_{8\_4\_4} \numberthis 
   \end{align*}
  where $ 8\_4\_4 $ is the baseline configuration, $ 8\_2\_4$ is the DL1 training configuration, $ 8\_4\_3$ is the IL1 training configuration and $1\_4\_4 $ is the L2 training configuration.\\
\emph{\bf{CPI model due to extra cache misses in lower-level caches ($CPI_{EM}$):}}
The most challenging aspect of the problem we solve is capturing the interplay between the different caches in a multi-cache way-disabling configuration.
Particularly, extra misses can occur in a multi-cache way-disabling configuration that do not occur either in the baseline or in the training single-cache way-disabling configurations. This is caused by the interactions between the higher-level and low-level caches. For example, this happens when an L2\$ hit in a single-cache training configuration becomes an L2\$ miss in the predicted multi-cache way-disabling configuration. Such interplay is not captured by the  $CPI_{L}$ model. Thus, we use $CPI_{EM}$ to encapsulate this behavior defined as follows:
This model is defined as follows:
\begin{equation}
C_{EM}= EM_{L2\$} * Penalty_{L2\$} 
\label{eq:extra}
\end{equation}
,where $EM_{L2\$}$ are the extra misses in L2\$ and $Penalty_{L2\$}$ is the cycle penalty per L2\$ miss. 
$EM_{L2\$}$ can be estimated using the following:
\begin{align}
EM_{L2\$}=TM_{L2\$}-M_{L2\$}  
\end{align}
,where the extra misses of L2\$ is the difference between the total expected L2\$ misses ($TM_{L2\$}$) and the L2\$ misses from the training configurations for $L2\$$ ($M_{L2\$}$).
More specifically, $TM_{L2\$}$ is estimated by:
\begin{align} 
TM_{L2\$}=M_{L2\${\rightarrow}D\$} +M_{L2\${\rightarrow}I\$} 
\label{eq:6}
\end{align}
Equation (\ref{eq:6}), aims to capture the total number of L2\$ misses of predicted configuration that mainly stem from two sources: 1) L2\$ misses caused by DL1\$ cache misses that access L2\$ $(L2\${\rightarrow}D\$)$ and 2) L2\$ misses caused from IL1\$ cache misses that access L2\$ $(L2\${\rightarrow}I\$)$. 
To do so,  $M_{L2\${\rightarrow}D\$}$ and $M_{L2\${\rightarrow}I\$}$ are calculated as follows: 
\begin{equation}
M_{L2\${\rightarrow}D\$}= M_{D\$} *MR_{L2\${\rightarrow} D\$}
\end{equation}
,where $M_{D\$}$ are the total misses from DL1\$ and, $MR_{L2\${\rightarrow} D\$}$ is the miss rate of the DL1\$ accesses in L2\$.
\begin{align}
M_{L2\${\rightarrow}I\$}= M_{I\$} *MR_{L2\${\rightarrow} I\$}
\end{align}
where $M_{I\$}$ are the total misses from IL1\$, and $MR_{L2\${\rightarrow} I\$}$ is the miss rate of the IL1\$ accesses in L2\$.

\begin{table}
\centering 
\caption{Processor Configuration}
\vspace{-0.25cm}
\begin{scriptsize}
\begin{tabular}{|l|l|} \hline
Parameter description   &Setting \\ \hline \hline
Pipeline depth      &15 stages \\
Fetch/Decode/Issue/
Commit&Up to 4/4/6/4 instructions\ per cycle \\
Line Predictor      & 4096 entries \\
RAS         &16 entries \\
Indirect Jump Predictor         &512 entries \\
Branch Predictor    &8$\:$KB gshare \\
Branch Resolution   &In-order\\
Prefetching & Disabled\\
Issue Queue/Reorder buffer       &40 INT entries, 20 FP entries / 128 entries \\
L1 instruction cache (IL1\$) &  4-way, 64$\:$B blocks, LRU \\
L1 data cache (DL1\$)& 4-way, 64$\:$B blocks, LRU\\
L2 unified cache (L2\$)& 8-way, 64$\:$B blocks, LRU\\ 
\hline
\end{tabular}
\end{scriptsize}
\vspace{-0.37cm}
\label{tab:configuration}

\end{table}

\begin{table} [t]
\caption{Execution time in hours for the exhaustive simulation-based and INTERPLAY approaches}
\centering \linespread{1}
\begin{scriptsize}
\centering
\begin{tabular}{|l|c|c|} \hline
Benchmark &Simulation-Based Approach   &INTERPLAY \\ \hline \hline
astar	&52.60	&5.75 \\
bwaves	&16.00	&1.75 \\
bzip2	&16.28	&1.78 \\
cactusADM	&13.25	&1.45 \\
gamess	&31.30	&3.42 \\
gcc	&44.93	&4.91 \\
GemsFDTD	&27.02	&2.95 \\
gobmk	&43.55	&4.76 \\
gromacs	&16.02	&1.75 \\
lbm	&18.33	&2.01 \\
leslie3d	&12.50	&1.37 \\
libquantum	&26.48	&2.89 \\
mcf	&37.80	&4.13 \\
milc	&25.67	&2.81 \\
namd	&47.52	&5.19 \\
omnetpp	&44.78	&4.89 \\
perlbench	&33.12	&3.62 \\
sjeng	&25.23	&2.76 \\
soplex	&47.95	&5.24 \\
sphinx3	&54.53	&5.96 \\
zeusmp	&19.98	&2.19 \\\hline
Total	&654.85	&71.62  \\

\hline
\end{tabular}
\end{scriptsize}

\label{tab:CPUTime}
\vspace{-0.7cm}
\end{table}

\begin{figure*}[t]
        \centering
        \begin{subfigure}{0.47\textwidth}  
            \centering 
            \includegraphics[width=\textwidth]{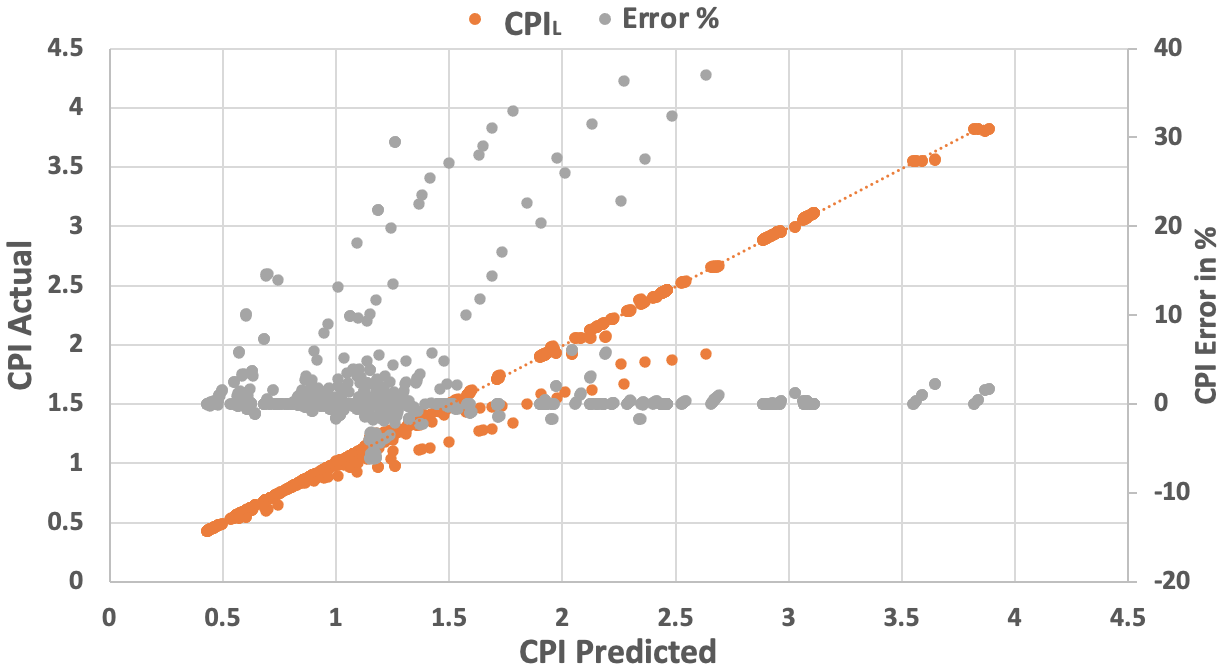}
            \caption[]%
            {{\small Predicted CPI Linear $(CPI_{L})$}}    
            \label{fig:CPIIdep}
        \end{subfigure}
        \begin{subfigure}{0.48\textwidth}  
            \centering 
            \includegraphics[width=\textwidth]{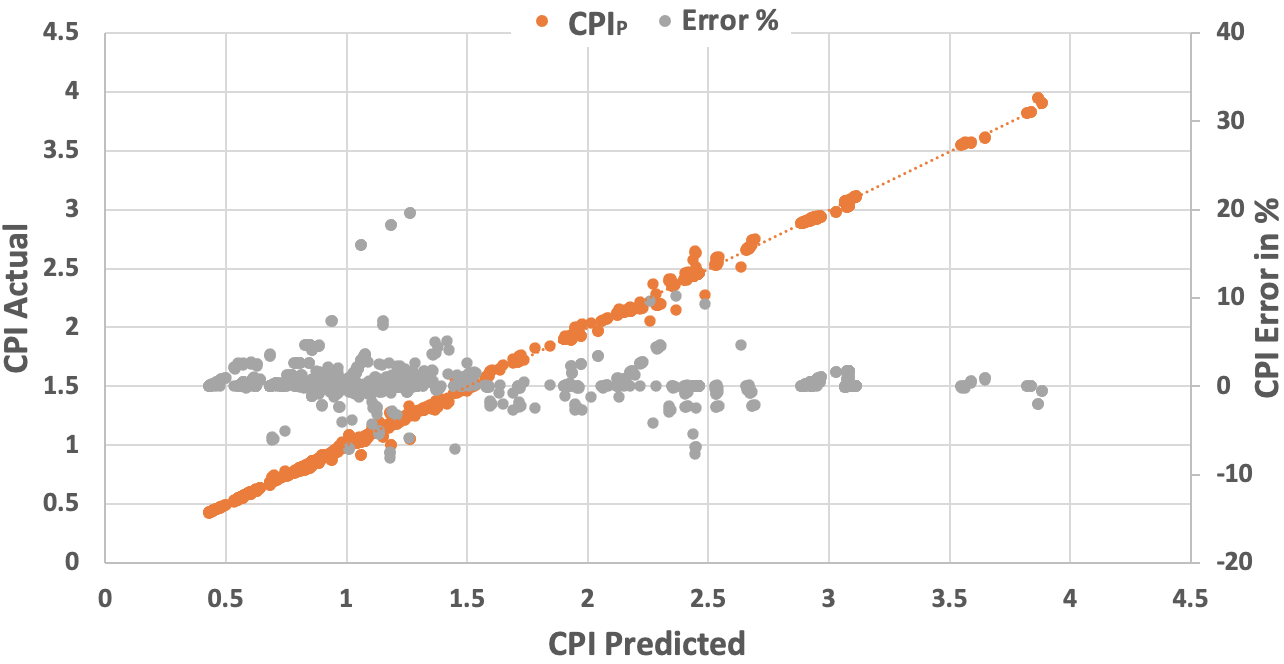}
            \caption[]%
            {{\small Predicted $CPI_{P}= (CPI_{L} + CPI_{EM})$}}    
            \label{fig:CPIdep}
        \end{subfigure}
        \hfill
        \vspace{-0.2cm}
        \caption[ ]
        {\small Cycles per Instruction (CPI) for model-based prediction for (a) $CPI_{L}$\   and\ (b) $CPI_{P}$, for a 2MB L2\$ cache} 
        \label{fig:CPIres}
        \vspace{-0.7cm}
    \end{figure*}

\begin{figure}[t]
    \centering
    \includegraphics[scale=0.29]{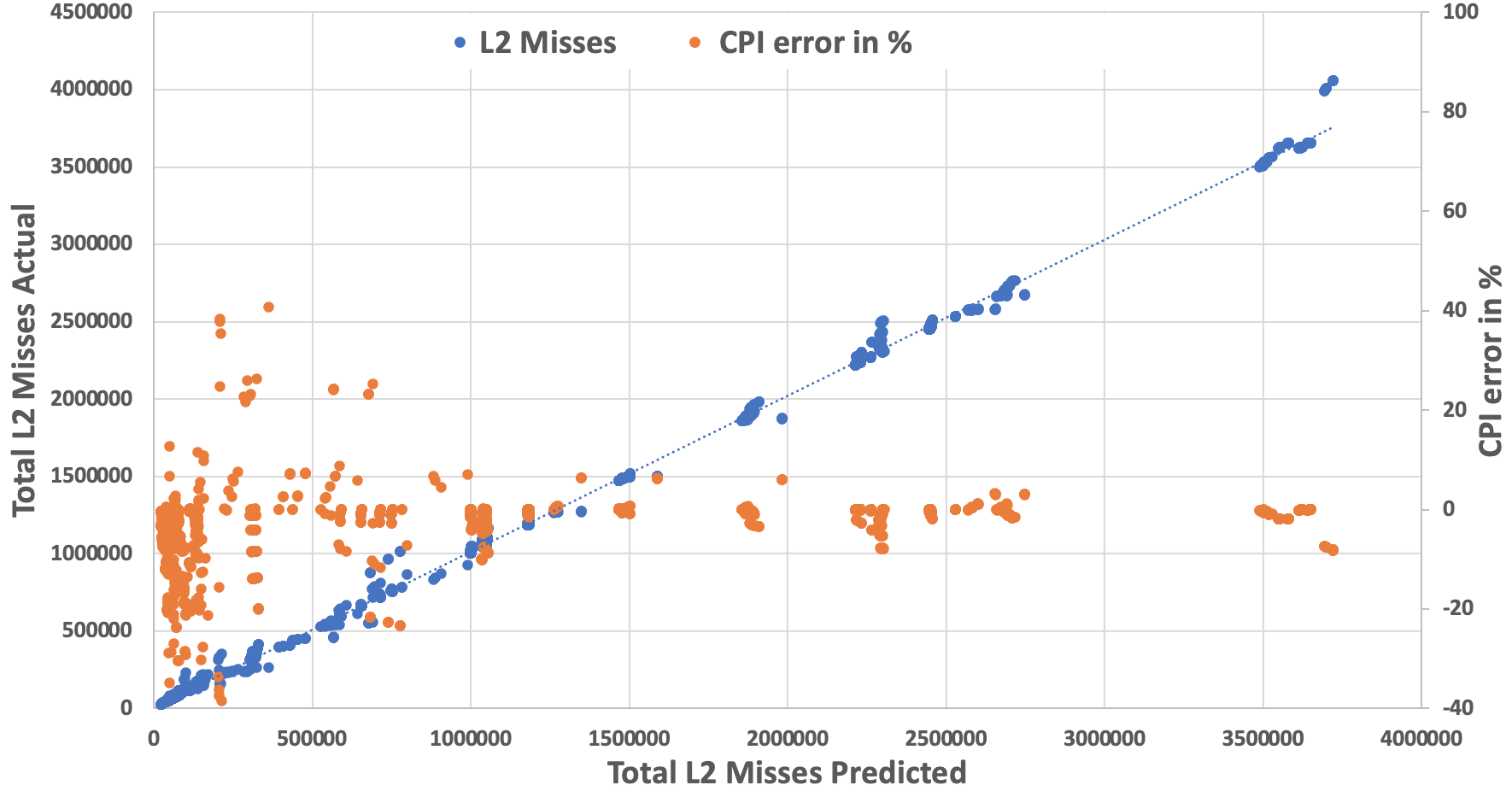}
    \caption{Predicted $L2\$$ Misses using $TM_{L2\$}$}
    \label{fig:L2Misses}
    \vspace{-0.7cm}
\end{figure}
The L2\$ miss rates of the predicted configurations are determined from the training data provided by the simulated single-cache configuration as the ratio of the number of L2\$ accesses, due to higher-level cache misses, that cause miss in L2\$, over the total number of L2\$ cache accesses due to the higher-level cache misses. Thus, the two miss rates are calculated using the following:
\begin{align*}
MR_{L2\${\rightarrow} D\$}=\begin{dcases*} M_{L2\${\rightarrow}D\$,L2T}/M_{D\$,L2T} &, \textit{(i)}\\
M_{L2\${\rightarrow}D\$,DT}/M_{D\$,DT} & , \textit{(ii)} \numberthis
 \label{miss1}
\end{dcases*}
\end{align*}
,where \textit{(i)} $M_{L2\${\rightarrow}D\$,L2T}\geq M_{L2\${\rightarrow}D\$,DT}$
and \textit{(ii)} $M_{L2\${\rightarrow}D\$,L2T}<M_{L2\${\rightarrow}D\$,DT}$.
\begin{align*}
MR_{L2\${\rightarrow}I\$}=\begin{dcases*} M_{L2\${\rightarrow}I\$,L2T}/M_{I\$,L2T}&, \textit{(iii)} \\
M_{L2\${\rightarrow}I\$,IT}/M_{I\$,IT} &, \textit{(iv)}  \numberthis
  \label{miss2}
\end{dcases*}
\end{align*}
,where \textit{(iii)} $M_{L2\${\rightarrow}I\$,L2T}\geq M_{L2\${\rightarrow}I\$,IT}$
and \textit{(iv)} $M_{L2\${\rightarrow}I\$,L2T}<M_{L2\${\rightarrow}D\$,IT}$.

Conditions \textit{(i)-(iv)} help determine the training configuration statistics that will be selected to compute the predicted misses. 


The $TM_{L2\$}$ includes the additional misses that are already captured by the linear model $(CPI_{L})$ so we need to subtract them. These misses correspond to $M_{L2\$}$ which is computed based on the misses from the training configurations (DT, IT and L2T) as follows:
\begin{align*} 
M_{L2\$}&=(M_{L2\$,DT}-M_{L2\$,B})+(M_{L2\$,IT}-M_{L2\$,B})+\\
&(M_{L2\$,L2T}-M_{L2\$,B})\numberthis
\end{align*}


Finally, to estimate the miss penalty in cycles per L2\$ cache miss, needed in eq (\ref{eq:extra}), we use the following:
\begin{equation}
Penalty_{L2}= (C_{L2T}-C_{B})/M_{L2\$,L2T} 
\end{equation}
,which determines the per miss cycles in the L2\$ and represents the difference in clock cycles between the baseline and L2\$ training configurations, divided by the misses in L2\$ in the training configuration L2T. 
\section{Experimental Setup} \label{setup}
\vspace{-0.17cm}


The simulation experiments in this work were performed using the cycle accurate simulator sim-alpha for ALPHA processor \cite{desikan2001sim}. The simulator is extended to support all the combinations of way disabling for all the cache levels (DL1\$, IL1\$, L2\$) leaving at least one operational way in each cache, and to monitor all the related performance statistics.
The key parameters of the simulated processor conﬁguration are summarized in Table \ref{tab:configuration}. The experiments are conducted for 100M committed instructions of 21 SPEC CPU2006 benchmarks. An in-house SimPoint \cite{sherwood2002automatically}-like tool is used to select the regions to simulate.

Two types of experimental results are reported: simulation and analytical based. The exhaustive simulation-based results are used to validate the accuracy of the proposed model. The validation compares the values obtained by simulations against the values predicted by the proposed model. 
The first set of results are for a processor with a 2MB L2 cache. To analyze further the model's accuracy and further highlight its generality, we have also experimented with a smaller L2 cache size of 256KB L2 that is still 8-way. 
For the training dataset we use a set of 14 single-cache way-disabling configurations per benchmark and for the prediction dataset we use the set of all possible 114 multi-cache way-disabling configurations.

\section{Results} \label{result}

\subsection{Execution time comparison of naive exhaustive simulation-based versus INTERPLAY}
We first analyze the execution time for all 21 benchmarks for the two approaches, exhaustive simulation vs combined simulation and model based approach (INTERPLAY). The results are obtained when running the simulations on a Xeon server and are presented in Table \ref{tab:CPUTime} for each benchmark. The second and third column represent the execution time in hours for simulation-based approach and INTERPLAY approach, respectively. As it can be seen in Table \ref{tab:CPUTime}, INTERPLAY approach is around 9.2 times faster than the simulation approach. It is useful to note that this value matches the ratio 128/14 which corresponds to the total number of multi-cache way-disabling configurations over the INTERPLAY training configurations. 

\begin{figure*}[t]
        \centering
        \begin{subfigure}{0.45\textwidth}  
            \centering 
            \includegraphics[width=\textwidth]{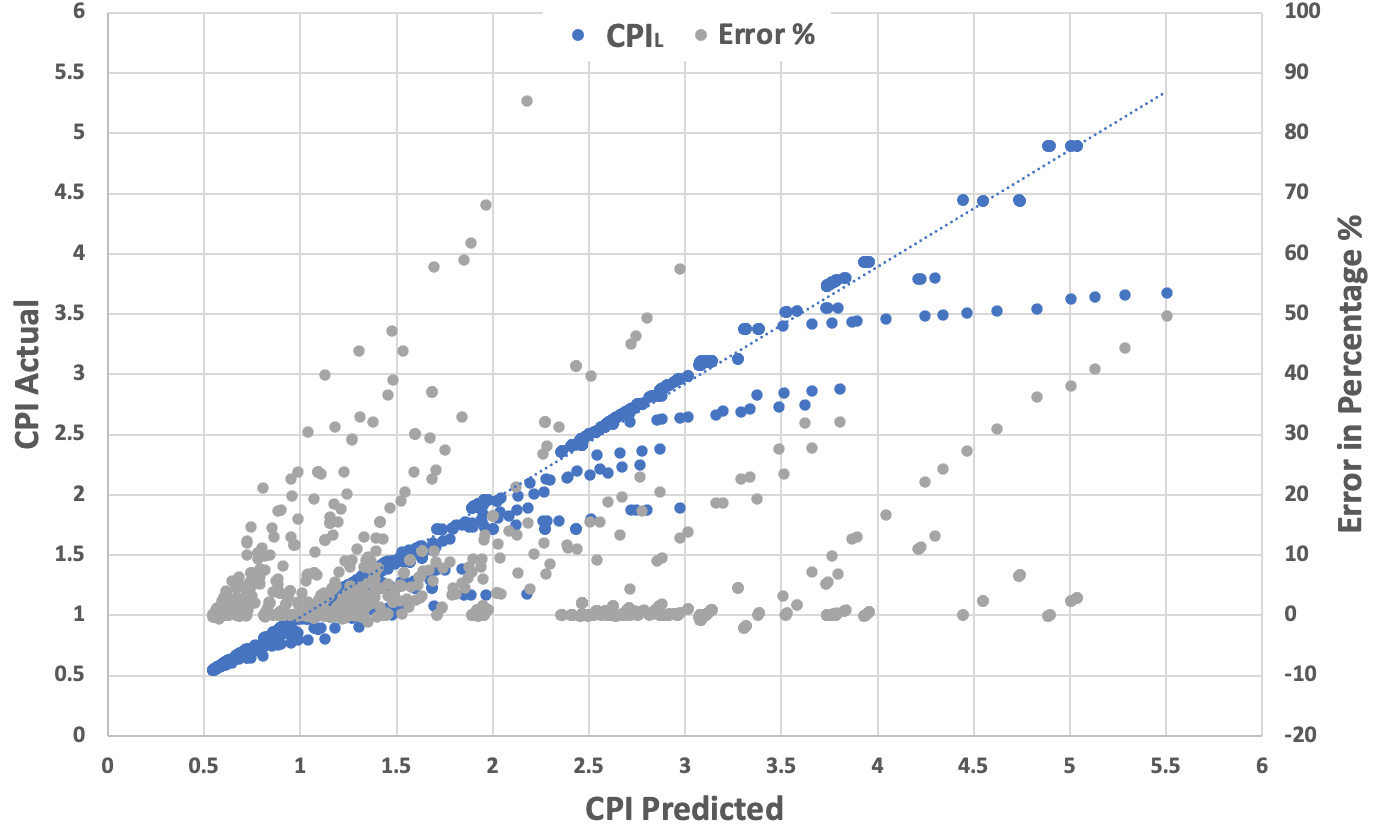}
            \caption[]%
            {{\small Predicted CPI Linear $(CPI_{L})$} \vspace{-0.2cm}}    
            \label{fig:CPIIdep}
        \end{subfigure}
        \begin{subfigure}{0.46\textwidth}  
            \centering 
            \includegraphics[width=\textwidth]{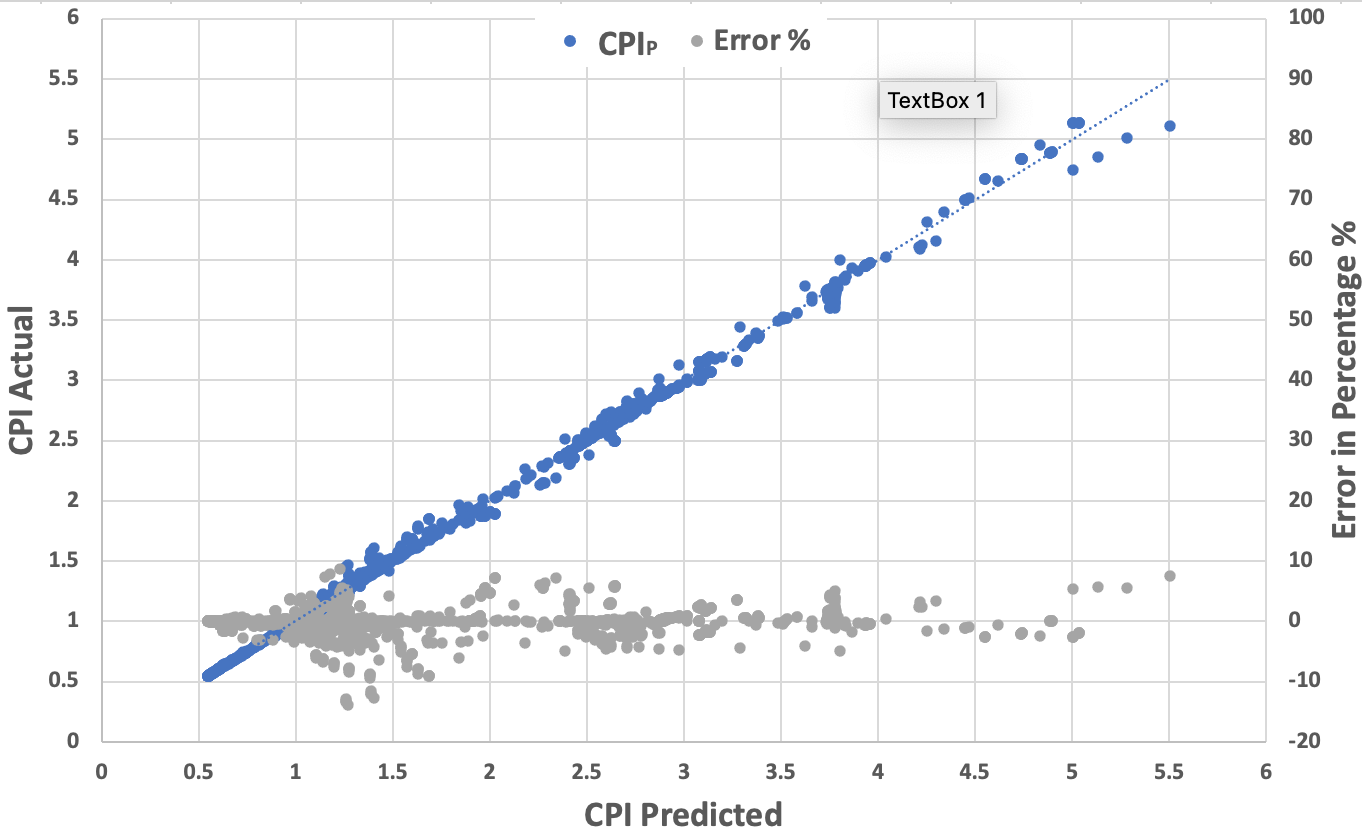}
            \caption[]%
           {{\small Predicted $CPI_{P}= (CPI_{L} + CPI_{EM})$} \vspace{-0.2cm}}    
            \label{fig:CPIdep}
        \end{subfigure}
        \hfill
        \caption[ ]
        {\small Cycles per Instruction (CPI) for model-based prediction for (a) $CPI_{L}$ and\ (b) $CPI_{P}$, for a 256KB L2\$ cache} 
        \label{fig:CPIres256}
        \vspace{-0.7cm}
    \end{figure*}

\subsection{Model-Based Approach Accuracy Results}
We evaluate the accuracy of the proposed predicting model for $CPI_{P}$, that includes both $CPI_{L}$ and $CPI_{EM}$. We also present results for $CPI_{L}$ alone, to highlight the need for both models. 
For these results we use CPI and we present the accuracy of the model in terms of percentage error. The error quantifies the difference of the proposed model with the actual exhaustive simulation results. Figures \ref{fig:CPIres}(a)-(b) show the CPI predicted (x-axis) versus the actual CPI (y-axis), as well as, the error in percentage (secondary y-axis) for $CPI_{L}$ and $CPI_{P}$, respectively, per benchmark and cache way-disabling configuration. 
In order to quantify the accuracy of our model we set-up a 5\% threshold as a maximum permissible error. Thus, error values that are above +5\% or below -5\% are considered as failed predictions. Consequently, the accuracy of $CPI_{L}$ is 96.42\% with an average error of 10\% and a maximum error of 37\%.
On the other hand, considering the $CPI_{P}$ model, as shown in Figure \ref{fig:CPIres}(b), we can see a prediction error decrease, giving an accuracy of 98.40\% with an average error of 3\% and a maximum error of 19\%.

The accuracy of the performance prediction is correlated with the accuracy that L2\$ misses are predicted, depicted in Figure~\ref{fig:L2Misses} which shows the predicted L2\$ misses (x-axis) estimated using eq. 7, the actual L2\$ misses (y-axis) and the error on the secondary y-axis. Figure \ref{fig:L2Misses} clearly demonstrates that in most of the cases where we experience a larger prediction error, the actual number of L2\$ misses is relatively small (see 0-100K misses in Figure~\ref{fig:L2Misses}), which means that the overall impact on performance will be negligible. 
Furthermore, we investigated the prediction accuracy using different thresholds and the results show an accuracy of 92.96\% with 2\% error threshold and 88.09\% with 1\% error threshold for the $CPI_{P}$ model versus 89.88\% with 2\% error threshold and 81.13\% with 1\% error threshold for the $CPI_{L}$ model. This clearly demonstrates the need for the $CPI_{P}$ model, which considers the interplay between lower-level and higher-level caches.

To further demonstrate the effectiveness of the model we also analyze INTERPLAY with smaller L2\$ cache which can cause a higher number of L2\$ misses.
Figures \ref{fig:CPIres256}(a)-(b) show the CPI prediction with a 256KB L2\$ for the $CPI_{L}$ model and the $CPI_{P}$ models, respectively. As it can be seen in Figure \ref{fig:CPIres256}(a), the prediction error in model $CPI_{L}$ increases considerably in the case of the smaller cache (compared with Figure \ref{fig:CPIres}(a)). However, when model $CPI_P$ is used (Figure \ref{fig:CPIres256}(b)), the prediction error remains small, similar to that in the larger cache (Figure \ref{fig:CPIres}(b)). In particular, $CPI_{P}$ for the smaller cache provides an accuracy of 96.8\%.

Finally, we investigated all the cases that have a prediction error above $|5\%|$ and we determined that the $CPI_{P}$ fails when extra L2\$ misses cause other misses. We call these secondary misses, and by analyzing them a bit further, we found out that when a benchmark has high reusesness on specific cache blocks and these cache blocks are replaced regularly by other blocks due to the small capacity of the cache (due to cache-way disabling), then this can result in secondary misses on the frequently accessed blocks, that our current approach cannot predict.   

\section{Conclusions and Future work} \label{conclusion}
This paper proposes INTERPLAY framework that uses a small set of single-cache way-disabling simulations as training configurations for an analytical model to predict the performance of multi-cache way-disabling configurations. Some key novelty of the work is the intelligent selection of the training configurations that consist of single-cache way disabling configurations and a model that predicts the extra misses caused by the interaction of higher and lower level caches when both caches, have some ways disabled.
The framework has several uses. One of them is to use it at design time to help designers leverage trade-offs between performance degradation and number of field returns. 
Future work will consider three-level cache hierarchy and multicore processors.
Finally, it would be also interesting to apply and validate the proposed model at the cache disabling granularity of a cache line.

\end{spacing}
\vspace{-0.2cm}
\begin{spacing}{0.83}
\bibliographystyle{IEEEtran}
\bibliography{ref}
\end{spacing}


\end{document}